\documentclass{article}
\usepackage[T1]{fontenc}
\usepackage{cite}
\usepackage{amsmath,amssymb,amsfonts}
\usepackage{algorithm}
\usepackage{graphicx}
\usepackage{subcaption}

\usepackage{textcomp}
\usepackage{xcolor}
\usepackage{array}
\usepackage{hyperref}
\usepackage{booktabs}
\usepackage{balance}
\usepackage{algpseudocode}
\usepackage{perpage}
\date{}

\begin{document}

\title{Reinforcement Learning - based  Adaptation and Scheduling Methods for Multi-source DASH}

\author{
Nghia T. Nguyen, Long Luu, Phuong L. Vo, \\
Thi Thanh Sang Nguyen, Cuong T. Do, Ngoc-thanh Nguyen
\thanks{Nghia T. Nguyen, Phuong L. Vo, Thi Thanh Sang Nguyen, and Long Luu are with the International University - Vietnam National University, Ho Chi Minh City, Vietnam; Cuong T. Do is with Kyung Hee University, 446-701, Korea; Ngoc-thanh Nguyen is with Wroclaw University of Science and Technology, Poland.
\\
\indent Email: \{ntnghia, vtlphuong, nttsang\}@hcmiu.edu.vn, ITITIU18079@student.hcmiu.edu.vn, dothecuong@gmail.com, ngoc-thanh.nguyen@pwr.edu.pl.
\\
\indent This is an extended version of the conference paper \cite{icccipaper}.
This research is funded by Vietnam National University HoChiMinh City (VNU-HCM) under grant number DS2020-28-01. Dr. Phuong L. Vo is the corresponding author.}
}

\maketitle

\begin{abstract}
Dynamic adaptive streaming over HTTP (DASH) has been widely used in video streaming recently. In DASH, the client downloads video chunks in order from a server. The rate adaptation function at the video client enhances the user’s quality-of-experience (QoE) by choosing a suitable quality level for each video chunk to download based on the network condition. 

Today networks such as content delivery networks, edge caching networks, content-centric networks, \textit{etc.} usually replicate video contents on multiple cache nodes. We study video streaming from multiple sources in this work. In multi-source streaming, video chunks may arrive out of order due to different conditions of the network paths. Hence, to guarantee a high QoE, the video client needs not only rate adaptation, but also chunk scheduling.

Reinforcement learning (RL) has emerged as the state-of-the-art control method in various fields in recent years. This paper proposes two algorithms for streaming from multiple sources: \textit{RL-based adaptation with greedy scheduling} (RLAGS) and \textit{RL-based adaptation and scheduling} (RLAS). We also build a  simulation environment for training and evaluating. The efficiency of the proposed algorithms is proved via extensive simulations with real-trace data.

\end{abstract}

\section{Introduction}
 A significant part of Internet traffic today is video streaming~\cite{Cisco23}. Dynamic adaptive streaming over HTTP (DASH) is a primary technique used to stream a video from a server to a video player. 
In DASH, videos are encoded in multiple quality levels.
Furthermore, videos are partitioned into video chunks. Each chunk contains media data in a short interval of playback time.
Video players (clients) request the chunks with suitable quality levels based on the current network condition~\cite{Stockhammer11, Sodagar11, Lederer12, Iso09}. 
The downloaded chunks are buffered in the client’s memory before being played. Buffer size is the total playing time of the wait-to-be-played chunks. 
When a new video chunk is successfully downloaded, the buffer size increases by a chunk length. When a chunk is played, the buffer size is decreased by the chunk length. 
The buffer size has an upper threshold level. When the buffer exceeds the threshold, the client will pause downloading a new chunk, wait for the buffer to decrease below the threshold, and then resume downloading. 
The client \textit{rebuffers} when the chunk will be played is not in the buffer. 
Rebuffering causes video freezes.

The rate adaptation function in video clients is essential in providing a high quality-of-experience (QoE) for the user. Various adaptation methods are proposed for DASH. The throughput-based adaptation method chooses the quality level for the next chunk such that the bitrate does not exceed the estimated throughput~\cite{Jiang12, dash}. The throughput is usually estimated by the mean or harmonic mean of several last requested chunks.
The buffer-based methods observe the buffer level to decide the encoding quality level~\cite{Spiteri16, Huang14}. 
Both the throughput-based method and BOLA, a buffer-based method~\cite{Spiteri16}, are employed in Dash.js reference client \cite{dash}.
Some methods combine these two approaches \cite{Li14}.

\begin{figure}[h]
	\centering
	\includegraphics[width=7.5cm]{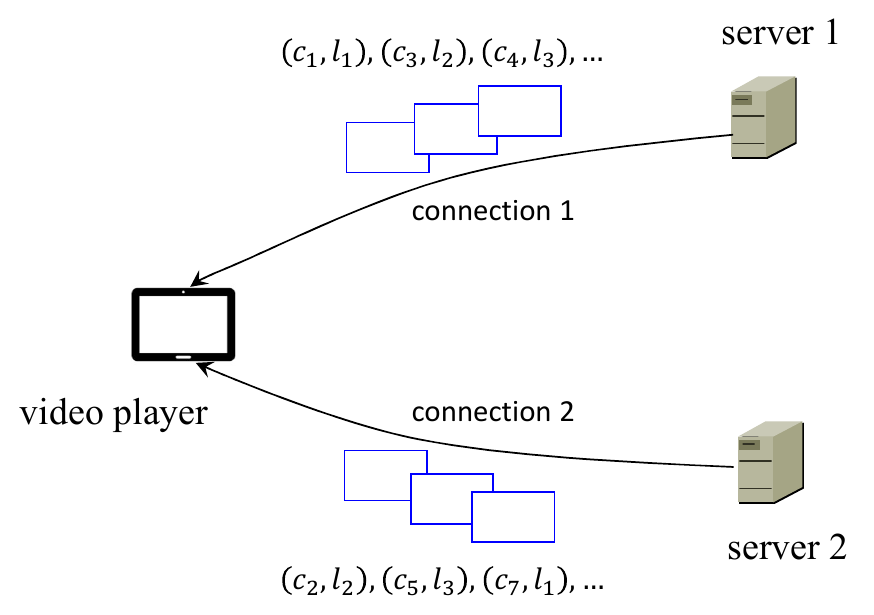}
	\caption{Multi-source video streaming.}
	\label{fig:multisource}	
\end{figure}

On the other hand, several networks today such as content delivery networks, edge caching networks, content-centric networks, \textit{etc.} replicate popular videos at the routers to reduce network congestion and delay. 
Utilizing multiple sources to stream a video to a user is studied in this paper. 
When streaming from multiple sources, quality control is much more complicated than streaming from a single source. 
In multi-source streaming, the quality control includes not only \textit{rate adaptation}, \textit{i.e.}, choosing the quality levels for the chunks, but also \textit{chunk scheduling}, \textit{i.e.}, which chunk indices are requested on each path (see Fig. \ref{fig:multisource}).
Due to the difference in the network conditions of the connections, the chunks may arrive at the video client out of order.
For example, assume that the maximum buffer size of the client is 3 chunks and there are two paths. With bad scheduling, path 2 is downloading chunk 2 while path 1, with very high throughput, already downloaded chunks 1, 3, 4. Therefore, the buffer is full, however, the video playing is frozen since the client waits for chunk 2.

Some previous works have studied multi-source streaming \cite{Chen16, Nikravesh19, Bentaleb20}. 
In \cite{Chen16}, MSPlayer can download video content from multiple servers. The authors in \cite{Chen16} consider the chunks with only one quality level, however, the chunk size varies. They focus on the chunk scheduling problem. The client estimates the path quality to request chunk indices and chunk sizes for the paths. 
In work \cite{Nikravesh19}, MP-H2 protocol is designed on top of HTTP/2. MP-H2 splits the video into many chunks, and the client requests chunks over multiple network connections such as wi-fi and cellular.
Chunk sizes are calculated based on bandwidth and round-trip-time of the connections. A chunk scheduling algorithm is applied to download the chunks over multiple paths without any adaptation method \cite{Nikravesh19}.
The work \cite{Bentaleb20} has proposed a bitrate adaptation algorithm for DASH, called DQ-DASH, that allows to download multiple video chunks from various servers in parallel to enhance QoE. Distributed queueing theory is applied to address the situation when multiple clients send requests to many servers simultaneously. Different from \cite{Chen16, Nikravesh19, Bentaleb20}, our proposed framework jointly apply both rate adaptation and chunk scheduling.

Reinforcement learning (RL) has been widely used in many fields recently. The works \cite{Claeys14, Mao17, Gadaleta17} have applied RL algorithms for single-source adaptive streaming. 
The actions of RL agents in these works are the quality levels of video chunks.
The work \cite{Claeys14} applied a Q-learning method for DASH. Buffer and network bandwidth are discretized for the discrete state space.
The study \cite{Mao17} used Asynchronous Advantage Actor-Critic (A3C) algorithm for rate adaptation.
The work \cite{Gadaleta17} has proposed D-DASH that applied a Deep Q-learning to choose the quality level for the chunks. 

In this work, we also use RL algorithm for rate adaptation and chunk scheduling in streaming from multiple sources. However, there are several challenges we cannot simply extend the RL framework for single-source streaming in \cite{Mao17, Gadaleta17} to multi-source streaming straightforwardly:
\begin{itemize}
\item The action space of multi-source streaming must be redesigned to integrate scheduling. An action must include a chunk index and a quality level.

\item The RL algorithms need a simulation environment to train the model. In the environment for single-source streaming, when the agent takes action, \textit{i.e.}, downloads a new chunk, the environment immediately returns a reward value associated with that chunk, which is calculated from the utility of the chunk’s quality, the quality switch penalty between two consecutive chunks and the rebuffering penalty. The rebuffering penalty is calculated when action is taken. However, in multi-source streaming, where the chunks may arrive to the client out of order, the simulation environment cannot estimate the rebuffering time right when an action is taken. 

\item The simulation environment for the single-source streaming is open~\cite{Mao17}. However, the simulation environment for the multi-source streaming is not available in the literature, as far as we know.
\end{itemize}

Multipath transmission control protocol (MPTCP) \cite{mptcp, mptcprfc, mreno} also utilizes multiple paths to transmit data from a source to a destination. It is shown that MPTCP achieves high throughput, provides a smooth hand-off, and improves the high availability of TCP connection. However, the MPTCP adoption has been prolonged because of the middlebox problem. Moreover, it requires modifying the kernels of both client and server. 
Our proposed framework can be applied to stream a video from a single source to a video player over multiple paths as MPTCP does.
However, the MPTCP is a source-control protocol at the transport layer, whereas our proposed multi-source streaming is a client-based control protocol at the application layer.
Therefore, the proposed multi-source streaming does not need to modify the kernel as well as overcomes the middlebox problem.

The contributions of our work include:
\begin{enumerate}
\item We propose two RL-based frameworks for rate adaptation and chunk scheduling in multi-source streaming called RL-based adaptation with greedy scheduling (RLAGS) and RL-based adaptation and scheduling (RLAS). 
\item We build an environment, which is an event-driven simulation that simulates a client downloading chunks from multiple sources and playing the chunks for training and testing.
\item We conduct extensive simulations with real-trace bandwidth to evaluate the performance of the proposed methods. 
Both RLAGS and RLAS outperform the other baseline methods used with greedy scheduling for multi-source streaming, \textit{i.e.}, throughput-based and BOLA.
The source code is available at \url{https://github.com/ntnghia1908/Master_Thesis}.
\end{enumerate}

This is the extended work of our conference paper \cite{icccipaper}. 
In this paper, RLAS improves the proposed algorithm in \cite{icccipaper} by using invalid action masking to avoid duplicate downloads. In addition, we propose RLAGS algorithm with a greedy scheduling. We also add more evaluations in various network scenarios.
The outline of the paper is as follows. Section I has presented the motivation and related works. Section II describes the RL model applied in rate adaptation and chunk scheduling for video streaming from multiple sources. The simulation environment and results are presented in Section III. Section IV concludes the work.

\section{Reinforcement learning frameworks for DASH}
This section describes the RL framework, including reward function, action space, and state space. Two chunk scheduling policies are considered, \textit{i.e.}, greedy and RL-based scheduling, which leads to two proposed algorithms, RLAGS and RLAS, respectively.

\subsection{Reward}

\begin{table}[h!]
	\centering
	\caption{Main notations}
	\begin{tabular}{c  p{6cm}}
	\toprule
	\textbf{Notations}	& \textbf{Descriptions}     	\\
	\toprule
	$B^{\max}$	&	maximum buffer size	(in seconds)\\
	\midrule
	$N$			& 	number of video chunks \\
	\midrule
	$L$ 		& number of quality levels in action space \\
	\midrule
	$W$			& number of chunks in action space\\
	\midrule
	$\mathcal{A}$ & action space\\
	\midrule
	$r_t$		& reward estimated at step $t$\\
	\midrule
	$s_t$		& environment state at step $t$\\
	 \midrule
	$q_i$ 		& utility of quality level $i$\\
	\midrule
	$\beta$		& quality-switch coefficient  \\
	\midrule
	$\gamma$	& rebuffering coefficient\\
	\bottomrule
\end{tabular}
\label{table:parameters}
\end{table}

We apply a similar reward function used in \cite{Mao17, Gadaleta17}, which captures utility, switch penalty, and rebuffering penalty.
Assume that a time step begins when the client requests a video chunk. The episode ends when the client finishes playing the video. 

\subsubsection{Reward for single-source streaming:}
Assume that step $t$ starts when the client requests for chunk $t$, the reward at step $t$ in single-source adaptive streaming is given by \cite{Mao17, Gadaleta17} (see Table~\ref{table:parameters} for the notation descriptions):
\begin{align}
r_t = q_t - \beta \mid q_t - q_{t-1} \mid - \gamma \phi_t - \delta [\max (0, B^{min} - B_{t})]^2, \quad t=2,\ldots,N,
\label{eq:singlereward}
\end{align}
where
\begin{itemize}
\item  $q_t$ is the utility corresponding to the quality level of chunk $t$;
\item $\mid q_t - q_{t-1}\mid$ penalties the difference in quality levels between two consecutive chunks;
\item $\phi_t$ is rebuffering time is seconds;
\item $[\max (0, B^{min} - B_{t})]^2$ is an optional penalty that is applied whenever the buffer level is below a threshold $B^{min}$. This term helps to reduce the risk of rebuffering.
\end{itemize}
If $d_t$, \textit{i.e.}, the download time of chunk $t$, is greater than remaining time in buffer, which is $B_t$, then rebuffering time $\phi_t$ is $d_t - B_t$, otherwise, there is no rebuffering. Hence, the rebuffering time associated with chunk $t$ in single-source streaming is given by the following formula
\begin{equation}
\phi_t = \max(0,d_t - B_t).
\label{eq:singlerebuffering}
\end{equation}

\subsubsection{Reward for multi-source streaming:}
Formula \eqref{eq:singlerebuffering} is not correct in the multi-source streaming environment since the buffer at the client may not store consecutive chunks. For example, the buffer may have chunks 3, 5, 6, and 7, while chunk $4$ has not fully received on the low-throughput path.
Therefore, in the multi-source environment, the reward is estimated when playing chunks in a step.
We assume that a step starts when the client requests a chunk and ends when the client requests a new chunk or reaches the end of the episode. The reward returned at step $t$ in the multi-source streaming environment is given by
\begin{align}
r_t = \sum_i q_i - \beta \sum_i \mid q_i - q_{i-1}\mid - \gamma \phi_t,
\label{eq:reward}
\end{align}
where $i$ is any chunk index played, and $\phi_t$ is the cumulative rebuffering time in step $t$.
The terms $\sum_i q_i$, $\beta \sum_i \mid q_i - q_{i-1}\mid$, and $\gamma \phi_t$ are called \textit{utility}, \textit{switching penalty}, and \textit{rebuffering penalty}, respectively.
 
\subsection{State space}
The state $s$ of the proposed reinforcement learning frameworks includes the following components 
\begin{itemize}
\item  vector of network throughput measurements of last 06 video chunks on each path;
\item vector of chunk sizes of $L$ quality levels of next $W$ chunks count from playing chunk (length $W \times L$);
\item the vector of quality levels of next $W$ chunks counted from the playing chunk, if the chunks have not yet downloaded, their quality level is set to 0;
\item current buffer size in seconds;
\item number of remaining chunks that have not yet played;
\item quality level of the playing chunk; and
\item download times of last 06 video chunks on each path.
\end{itemize}

\subsection{Scheduling policies and action spaces} 
We assume that the request for a new chunk is sent on a path right after the downloading chunk on that path is fully received if the buffer size is under $B^{\max}$. Otherwise, the client will pause sending a new request.
We consider two scheduling policies, \textit{i.e.}, greedy scheduling and RL-based scheduling, corresponding to two proposed methods, RLAGS and RLAS, respectively.

\subsubsection{Greedy scheduling}
In greedy scheduling, the chunk is requested in order. When the client downloads a new chunk from a source, it requests the chunk index, the smallest index that has not been or is being downloaded.
Therefore, RLAGS agent only decides the quality level of the chunk to request.
The action space of RLAGS includes the quality levels of video chunks:
\begin{equation}
    \mathcal{A}^{\rm{RLAGS}} = \{l_i|i=1,\ldots, L\},
\end{equation}
where $L$ is the number of quality levels of video.

For example, in Fig. \ref{fig:actionspace} (upper figure), chunks 1-3 have been played, chunk 4 is playing, and chunk 5 has already been requested. The next request is for chunk 6, with the quality level decided by RLAGS.

\subsubsection{RL-based scheduling}
RLAS method uses RL-based scheduling. When the client requests a new chunk, the RL agent decides both the index and quality level. 
Assuming that the maximum number of chunks stored in the video buffer is $W$, the number of quality levels is $L$.
Action space of RLAS method is defined as
\begin{equation}
    \mathcal{A}^{\rm{RLAS}} = \{(c_i, l_j) | c_i \in [1,W], j \in [1, L]\}.
\end{equation}
If the playing chunk is $c_t$ and the RL agent takes action $a_t=(c_i, l_j)$ to download chunk on a path at time step $t$, the client will download chunk index $c_t+c_i$ at quality $l_j$ on this path.

\begin{figure}[h]
\centering
	\includegraphics[width=9cm]{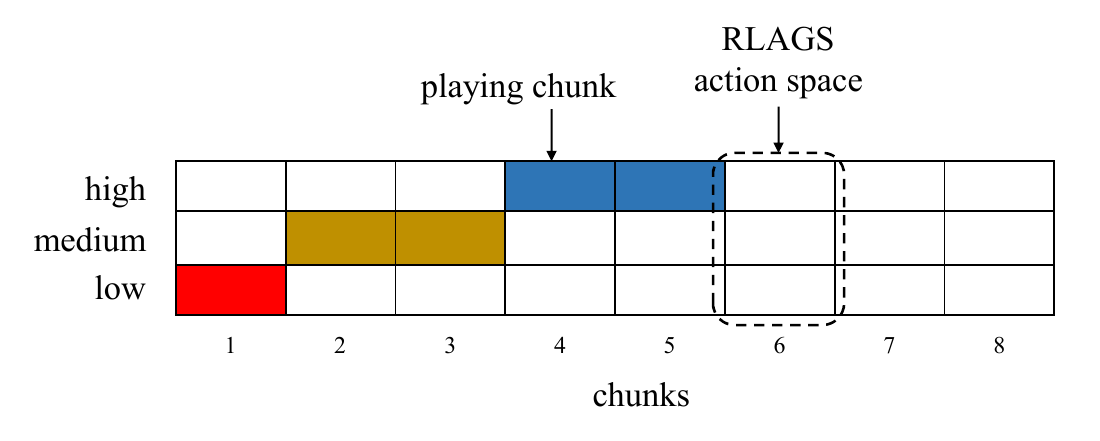}\\
	\includegraphics[width=9cm]{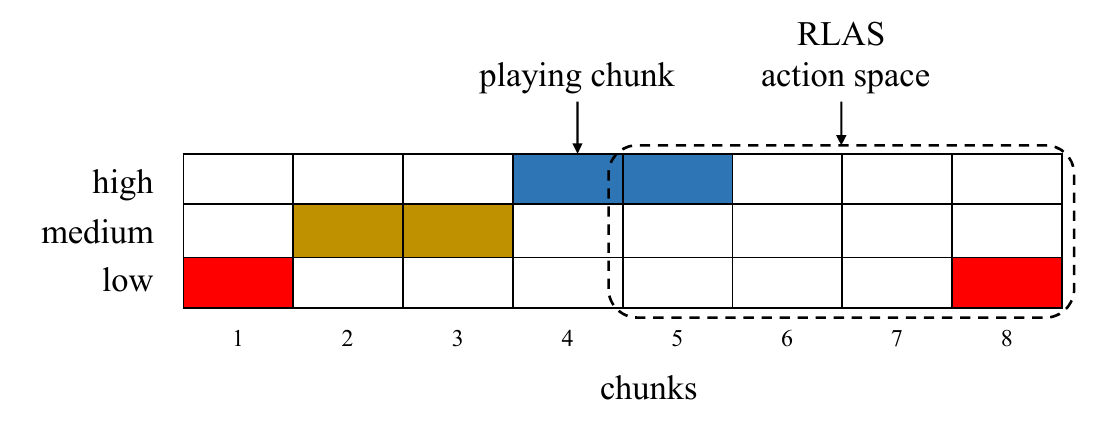}
	\caption{The action spaces of RLAGS and RLAS. (The shade regions represent the chunks that have been requested.)}
	\label{fig:actionspace}	
\end{figure}

For example, in Fig. \ref{fig:actionspace}, if quality levels for each chunk are \textit{low}, \textit{medium}, and \textit{high} ($L=3$), $W=4$. Assume that at current time $t$, the playing chunk is $c_t=4$ and the RL agent takes action $a_t = (3,2)$. It means that the agent will download chunk index $4+3=7$ with quality level $2$, which is \textit{medium} quality (see Fig. \ref{fig:actionspace}).

\subsubsection{Invalid action masking}
With RLAS, there are invalid actions in some steps.
Firstly, the two-dimension action space of RLAS allows the possibility of re-download the same chunk index again if that chunk has not been played. 
The chunk index already downloaded is considered an invalid action to avoid duplicate download.
Secondly, since RLAS's action space is a sliding window that shifts forward by one chunk when a new chunk is played, some actions are \textit{invalid} when the number of remaining chunks is less than the window side $W$. Hence, the valid actions of RLAS are given by
\begin{equation}
	\{(c_i, l_j) \mid c_i \in [1, \min{[W, N-c_t]}, c_i \textrm{ has not been requested}, j \in [1,L]\},
\end{equation}
where $N$ is the number of chunks of the video, and chunk $c_t$ is the chunk being played.

There are several approaches to dealing with invalid actions. Two common ones are \textit{invalid action penalty} and \textit{invalid action masking}. 
With the invalid action penalty approach, the rewards resulting from the invalid actions are set to negative values.
With invalid action masking, the action is sampled among the valid actions in each step.
These approaches are well investigated and implemented in \cite{invalidaction}. 
The authors show that the invalid action masking outperforms the other approaches theoretically and empirically with the state-of-the-art proximal policy optimization (PPO) algorithm \cite{invalidaction}.
Therefore we also apply PPO in our evaluations.

\subsection{PPO for multisource DASH}

PPO is a policy gradient algorithm that uses two networks, actor and critic, like A2C or A3C. The actor network estimates the policy directly from the state.  A baseline is subtracted from the return to reduce the variance of a policy gradient algorithm. A common-used baseline is the value function, which is estimated by a critic network. To accelerate the training, PPO algorithm could also use multiple copied environments in parallel, similar to A2C and A3C.

However, PPO is an improvement from A2C/A3C. To prevent the catastrophic drop in the performance of the traditional actor-critic algorithms, PPO constraints the change in policy between two consecutive training steps by introducing a new clipped surrogate objective. PPO has shown a reliable performance and is used in many RL applications. We can see the detail of PPO algorithm in \cite{ppo}.

We utilize Stable Baseline3 \cite{SB3} library to implement PPO in training and evaluation. Stable Baseline3 includes a set of reliable implementations of deep RL algorithms and is used in many applications. The invalid action masking function is also provided with PPO in Stable Baseline3.

\section{Evaluations}

\subsection{Event-driven environment}

We build an environment that simulates the streaming from two sources, emulating the practical scenarios, \textit{e.g.}, a cell phone uses Wi-Fi and 4G to connect to video servers, or a laptop connects via Ethernet and Wi-Fi simultaneously. 
Scenarios having more than two sources can be easily extended by modifying \textit{reset} function. The simulation environment follows Gym interface to be compatible with Stable Baseline3 \cite{SB3}.\footnote{The source code of the environment is available at \url{https://github.com/ntnghia1908/Master_Thesis/blob/main/RLAS/menv_baseline.py}.}

The environment emulates a client downloading chunks on two paths parallelly and playing the received chunks. 
An array-type buffer, which stores downloaded chunk indices, is maintained during an episode. When the client fully receives a chunk, the chunk index is appended to the buffer, and the buffer size increases by a chunk length.
The client plays the chunks stored in the buffer sequentially. If a chunk is played, that chunk index is removed from the buffer, and the buffer size increases by a chunk length.

There are four main events, \textit{i.e.}, DOWN, PAUSE, PLAY, REBUFFER.
Every event has a timestamp, and the program runs through the events iteratively in time order until the end of the episode. 
DOWN and PAUSE events are associated with a path index, whereas PLAY and REBUFFER events are not.
\begin{itemize}
\item A DOWN event simulates sending a request for a chunk, say $c_t$, on a path. When the program encounters a DOWN event at time t, at timestamp \textit{t + downtime}, where downtime is the time from sending the request for $c_t$ to fully receiving the chunk,  a new DOWN event associated with a new chunk is generated if the buffer size is less than $B^{\max}$. The index and the quality level of the new chunk are decided by RLAGS or RLAS methods. Otherwise, a PAUSE event is generated if the buffer size exceeds $B^{\max}$.

\item A PAUSE event simulates pausing the download on a path due to the buffer size exceeding $B^{\max}$. If a PAUSE event is encountered at time $t$, with timestamp $t+sample$, where sample is a short period (0.05 second in our program), a new PAUSE event is generated if the buffer size exceeds $B^{\max}$; otherwise, a new DOWN event associated with a new chunk is generated.

\item A PLAY event occurs when the client starts playing a chunk, say chunk $c_t$. 
After a PLAY event, at time $t + chunk\_length$, a new PLAY event associated with chunk $c_t+1$ is generated if this chunk is available in the buffer; otherwise, a REBUFFER event is generated.

\item A REBUFFER event occurs when the chunk going to be played is not in the buffer.
After a REBUFFER event, at time $t+sample$, a new PLAY event associated with chunk $c_t+1$ is generated if this chunk is fully received; otherwise, a REBUFFER event is generated.
\end{itemize}


\subsection{Simulation settings}
We evaluate RLAGS and RLAS with Big Bug Bunny video \cite{Lederer12}. There are seven quality levels $300, 700, 1200, 1500, 3000, 6000, 8000$ Kbps ($L=7$).
Assume that the maximum buffer size of the client is $B^{\max} = 30$ seconds and the video chunk length is 4 seconds. The number of chunks in the action space is $W = \lfloor 30/04 \rfloor = 7$, which means that if the agent is playing chunk $i$, the maximum chunk index stored in the buffer is $i+7$.
We train with the first 60 chunks, which results in 240 seconds per episode. 
Table \ref{table:simparameters} shows the parameters of the simulation environment.

Two real-trace datasets are used: a broadband dataset provided by US Federal Communications Commission (FCC) \cite{Fcc19} and a 4G LTE Dataset collected from two major Irish mobile operators \cite{Raca18}.

\begin{figure}[h]
	\centering
	\includegraphics[width=8cm]{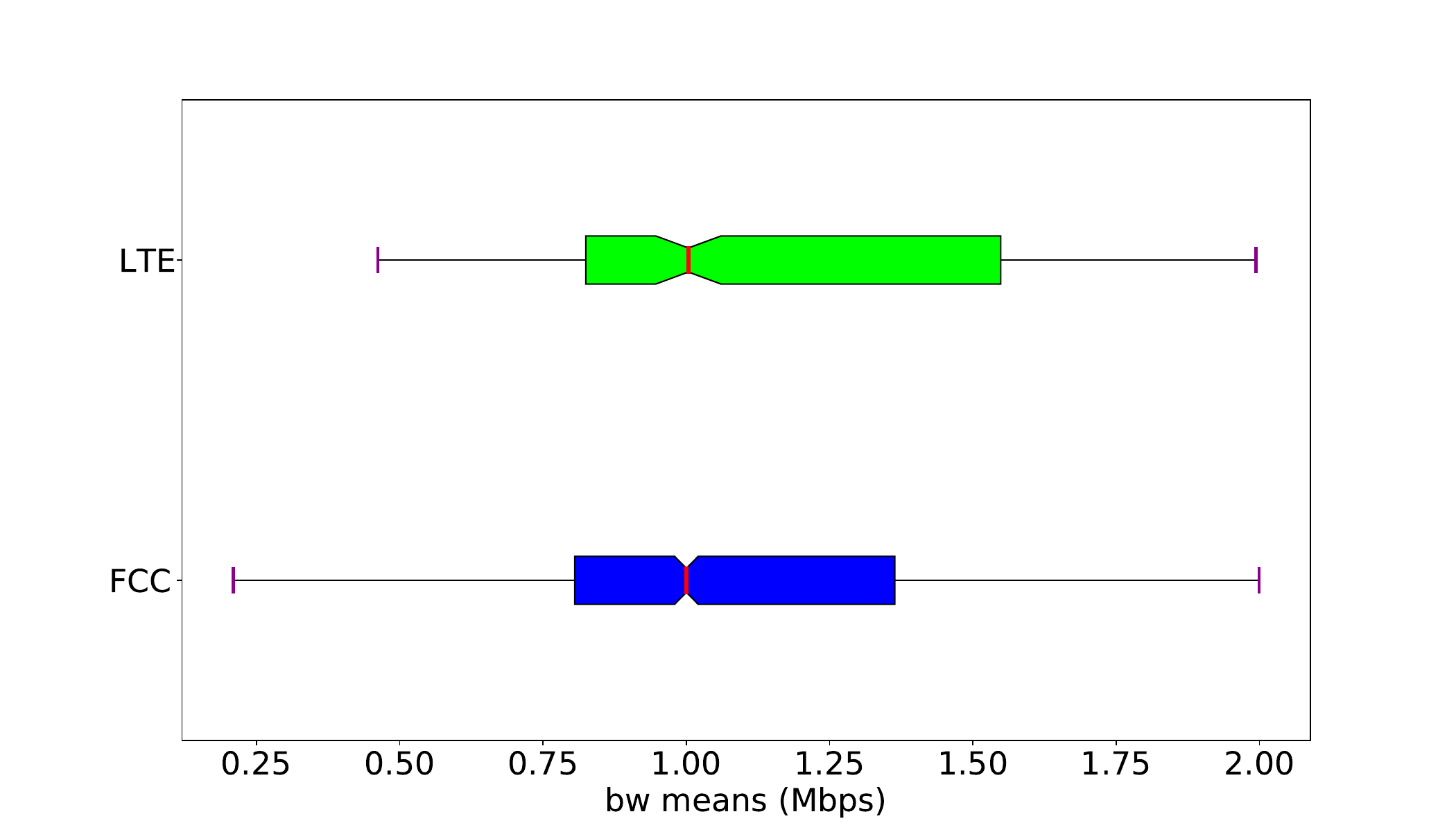}
	\caption{The distributions of the means of all traces of the datasets.}
	\label{fig:bandwidth-description}	
\end{figure}

The \textbf{FCC dataset} contains over one million throughput traces in the ``download speed'' category with a granularity of 10 seconds per sample \cite{Fccreport19}. (It is 5 seconds before 2016.)
The \textbf{4G dataset} has 135 traces, with around 15 minutes per trace, at 1-second granularity. The traces are collected from Irish mobile operators with five mobility patterns: static, pedestrian, car, bus, and train \cite{Raca18}. 

Since the real bitrates of 8,000 Kbps quality level of the almost chunks are less than 4000 Kbps \cite{Lederer12}, we choose the traces with the average throughputs in $[0.1, 2.0]$ Mbps, in which 1800 traces in the FCC dataset and 400 traces in the LTE one. 
Fig. \ref{fig:bandwidth-description} shows the distributions of the average throughputs of the traces from two datasets. 
We randomly select 80\% of traces in each dataset for training, the remaining ones are for testing.

\begin{table}[h!]
	\centering
	\caption{Simulation parameters}
	\begin{tabular}{p{3.5cm} c p{5.5cm}}
	\toprule	
	\textbf{Environment parameters}	& \textbf{Notations}     & \textbf{Values}	\\
	\toprule
	maximum buffer size & $B^{\max}$ & 30 seconds\\
	\midrule
	number of video chunks & $N$ & 60 chunks\\
	\midrule
	number of quality levels in action space & $L$ & 7 \\
	\midrule
	number of chunks in action space &  $W$ & 7\\
	 \midrule
	quality levels  & $l_i$& $[300, 700, 1200, 1500, 3000, 6000, 8000]$ Kbps\\
	\midrule
	utility & $q_i$& $\ln(\frac{l_i}{l_1})$\\
	\midrule
	quality-switch coefficient & $\beta$ & 1 \\
	\midrule
	rebuffering coefficient& $\gamma$ & 3.3 \\
	\bottomrule
\end{tabular}
\label{table:simparameters}
\end{table}

We compare RLAGS and RLAS methods with two well-known adaptation methods, \textit{i.e.}, throughput-based \cite{Stockhammer11} and BOLA \cite{Spiteri16} (a buffer-based) methods. 
These adaptations are originally designed for single-source video streaming. We apply greedy scheduling to extend them to multi-source streaming. 
In the throughput-based method, the quality of the next download chunk on one path is the highest quality level which is smaller than the harmonic mean of the last six chunks downloaded on that path.
 
Table~\ref{table:tunedparameters} lists some tuned hyper-parameters for RLAGS and RLAS. The not-listed hyperparameters are used with the default values provided by Stable Baseline3. 
We use fully connected neural networks with 64 nodes for each hidden layer. 
We tuned the number of hidden layers for the algorithms.
Round-trip-times of the network connections are uniformly random in $[50,100]$~ms.

Each proposed algorithm is trained in five runs, $30,000$ episodes each run.  Each episode chooses a random trace in the training set and starts at a random point. The throughput trace is circulated if the time from starting point of an episode to the end of the throughput trace is not enough for the time playing the episode.
The results are the average values in five runs.

\begin{table}[!h]
\centering
\caption{Tuned hyperparameters used in RLAGS and RLAS.}
\begin{tabular}{l p{3cm} p{1.5cm} p{1.5cm} p{1.5cm}}
\toprule
\textbf{Hyperparameters} & \textbf{Descriptions} & \textbf{Tuning ranges}  & \textbf{RLAGS} & \textbf{RLAS} \\ \toprule 
learning rate   & learning rate                                                 & uniform: [0.0001, 0.001]        & 0.000125   &   7.61e-05            \\ \midrule
batch size      & minibatch size                                                     & uniform: [59, 590]          & 411         &  530         \\ \midrule
n epochs        & number of epoch when optimizing the surrogate loss                 &  values: [10, 20, 30]         & 10          &  10         \\ \midrule
gamma           & reward  discount factor                                            & values: [0.99, 1.0]           & 0.99        &   1        \\ \midrule
gae lambda      & factor for trade-off of bias vs. variance for generalized advantage estimator  & values: [0.9, 0.95]  & 0.9         &   0.95        \\ \midrule
clip range      & clipping parameter                                                 & values: [0.2, 0.3]          & 0.3         &  0.2         \\ \midrule
vf coef         & value function coefficient for the loss calculation                & uniform: [0.2, 0.5]          & 0.317708    &  0.286954          \\ \midrule
ent coef        & entropy coefficient for the loss calculation                       & values: [0.0, 0.00001, 0.00000001]          & 0           &  0                       \\ \midrule

act func        & value function coefficient for the loss calculation                 & values: [128, 256, 512]       & 256    &  512  \\ \midrule
features dim    & value function coefficient for the loss calculation                 & values: [tanh, relu]           & relu   &  tanh\\ \midrule
policy net arch layer           & number of policy network layer                      & values: [1, 2, 3, 4]      & 1     &  3 \\ \midrule
policy net arch units           & policy network unit.                          & values: [64, 128, 256, 512] & 512    &  256\\ \midrule
value net arch layers           & number of value network layer                       & values: [1, 2, 3, 4]     & 3 & 4  \\ \midrule
value net arch units            & value network unit.                        & values: [64, 128, 256, 512] & 512  & 256\\ \bottomrule
\end{tabular}
\label{table:tunedparameters}
\end{table}

\subsection{Results}

\begin{figure}[h]
\centering
	\includegraphics[width=9.5cm]{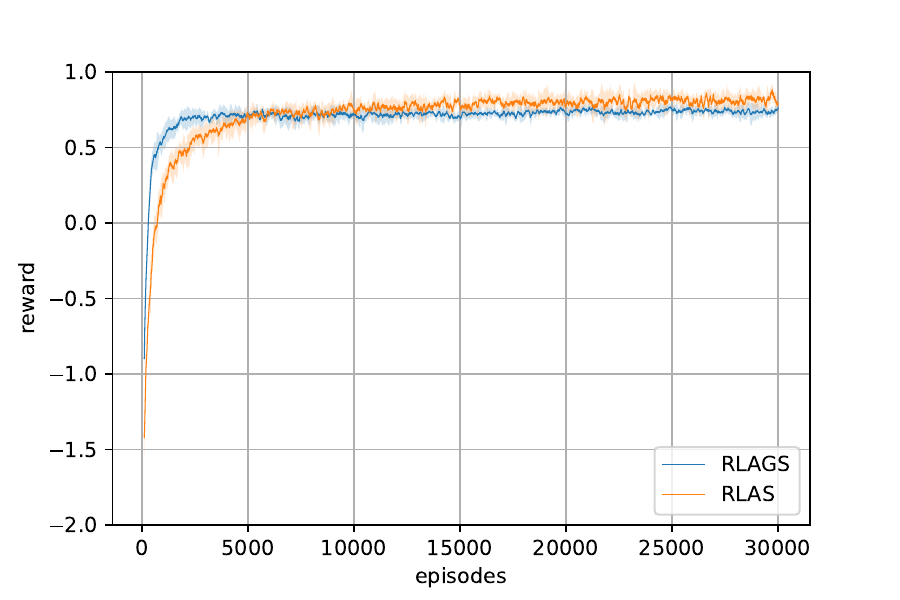}
	\caption{Convergence of training phases of RLAGS and RLAS methods. The lines and shadows are the means and the standard deviations of the running average rewards of five runs, respectively.}
	\label{fig:convergence}	
\end{figure}
Fig. \ref{fig:convergence} shows the convergence of both RLAGS and RLAS algorithms in training with turned parameters given in Table \ref{table:tunedparameters}. We can see that RLAGS converges faster than RLAS since RLAGS has fewer actions than RLAS, which are only quality levels. However, RLAS yields a higher average reward than RLAGS.

\begin{table}[!ht]
	\centering
	\caption{Rewards of ABR methods when one path is broadband and another path is LTE connection.}
	\begin{tabular}{p{1.2cm}cccc}
	\toprule
	\textbf{Methods} & \textbf{reward} &	\textbf{utility}	& \shortstack{\textbf{switch}\\\textbf{penalty}}	& \shortstack{\textbf{rebuffering}\\\textbf{penalty}}\\
\toprule
THGHPUT	&42.10	&68.56	&21.40	&5.06\\
\midrule
BOLA	&77.80	&129.75	&27.26	&24.70\\
\midrule
RLAGS	&88.35$\pm$2.04&97.86$\pm$2.29&8.53$\pm$1.01&0.98$\pm$0.52\\
\midrule
RLAS & \textbf{107.75$\pm$1.91} &130.68$\pm$5.87&17.51$\pm$4.88&5.42$\pm$2.62\\
\bottomrule
\end{tabular}
\label{table:fcc-lte}
\end{table}

We test the case when one path is a broadband connection, and the other path is an LTE connection. The reward, utility, switch penalty, and rebuffering penalty are given in Table \ref{table:fcc-lte}. The rewards yielded by RLAGS and RLAS are higher than the reward by throughput-based and BOLA methods. RLAS achieves the highest reward, and RLAGS results in a smaller rebuffering penalty.

\begin{figure}[h]
	\centering
	\includegraphics[width=9.5cm]{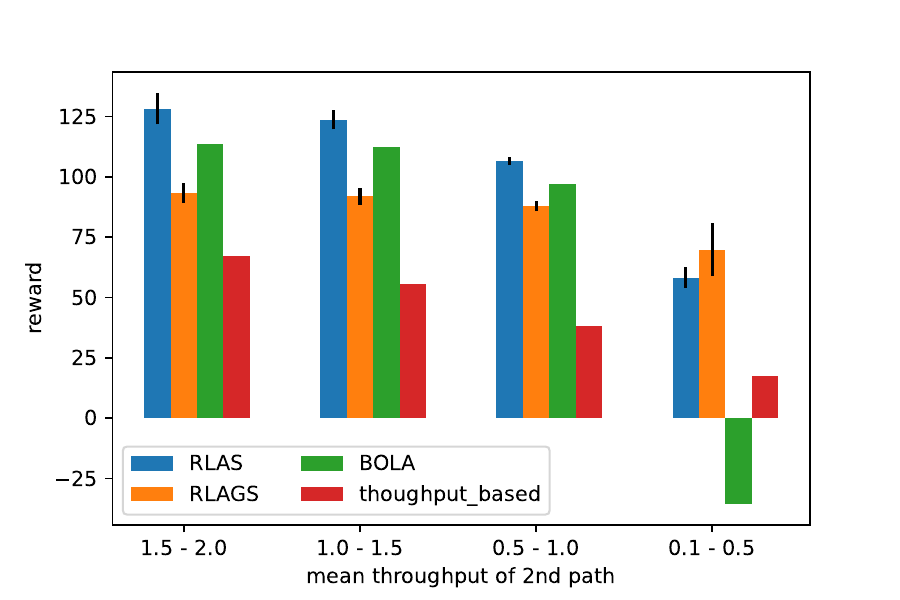}
	\caption{The test rewards of different adaptation methods when the mean bandwidth of the first path is in $[1.5, 2]$ Mbps and the mean bandwidth of the second path decreases gradually.}
	\label{fig:rewards}	
\end{figure}
We consider the performance of multisource streaming in the case when the difference between two paths increases gradually. Particularly, the mean bandwidth of the first path is from 1.5 Mbps to 2 Mbps and the mean bandwidth of the second path decreases gradually, in $[1.5, 2.0]$ Mbps, in $[1.0,1.5]$ Mbps, in $0.5, 1.0$ Mbps, and less than 0.5 Mbps.

We can see from Fig. \ref{fig:rewards} that the rewards of multisource streaming of all the methods decrease gradually. The rewards of RLAS are the highest in most of the cases, which shows the efficiency of the RL-based chunk scheduling.
BOLA yields a higher reward than RLAGS. However, in the extreme case when the mean bandwidth of two paths is very different, RLAGS and RLAS outperform the traditional methods.

\begin{table}[!ht]
	\centering
	\caption{Rewards of ABR methods with the mean bandwidths of the first path is in $[1.5,2.0]$ Mbps and of the second path is less than 0.5 Mbps.}
	\begin{tabular}{p{1.2cm}cccc}
	\toprule
	\textbf{Methods} & reward &	utility	& \shortstack{switch\\penalty}	& \shortstack{rebuffering\\penalty}\\
\toprule
THGHPUT	&17.34	&68.20	&32.03	&18.83\\
\hline
BOLA	&-35.78	&120.11	&19.10	&136.78\\
\midrule
RLAGS	&\textbf{66.61$\pm$4.94}	&92.68$\pm$2.66 &9.74$\pm$1.92	&16.33$\pm$7.69\\
\midrule
RLAS & 57.55$\pm$3.99	&105.69$\pm$1.77	&17.54$\pm$1.49	&30.59$\pm$3.76\\
\bottomrule
\end{tabular}
\label{table:rewards2005}
\end{table}

Table.~\ref{table:rewards2005} shows the performance of the adaptation methods in the extreme case: the average throughput of the first path is in $[1.5,2.0]$ Mbps, and of the second path is less than 0.5 Mbps.
Overall, RLAGS and RLAS outperform BOLA method, and their rewards are much higher than those of the throughput-based method. 
The reward of RLAGS is a bit higher than RLAS because it has fewer actions in the action space. Hence, the agent may be easier to learn the optimum.
The throughput-based method has the least rebuffering; however, it also has the lowest utility. 
BOLA has the highest utility but also the highest rebuffering penalty.
RLAGS balances the objectives: high utility, low number of switches, and small rebuffering penalties.

\begin{figure}[h]
	\centering
		\includegraphics[width=9cm]{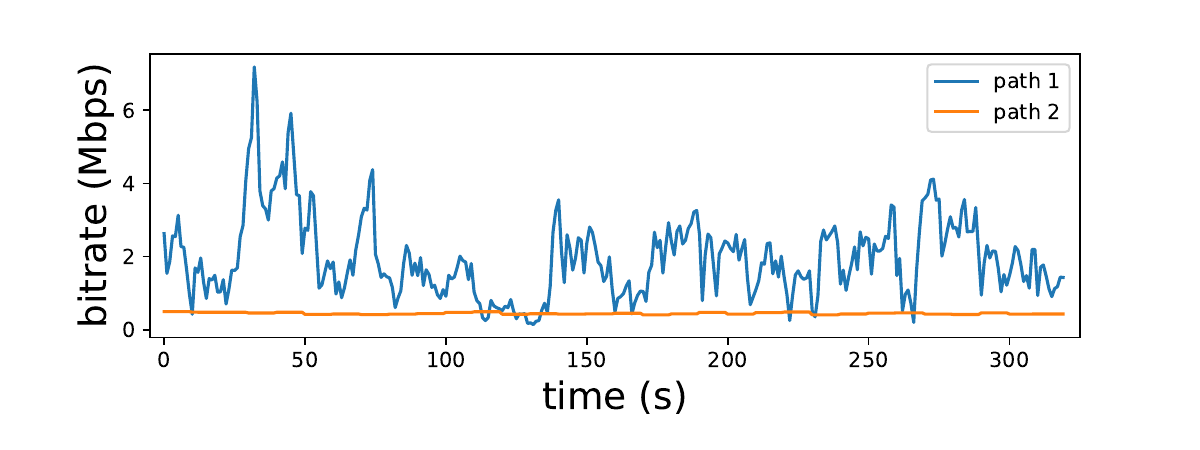}			
	\caption{A sample of throughput traces in the extreme case: the mean bandwidths of the first path is in $[1.5,2.0]$ Mbps and of the second path is less than 0.5 Mbps.}
	\label{fig:bwtraces}	
\end{figure}

\begin{figure}[ht]
	\centering
	\includegraphics[width=0.45\textwidth]{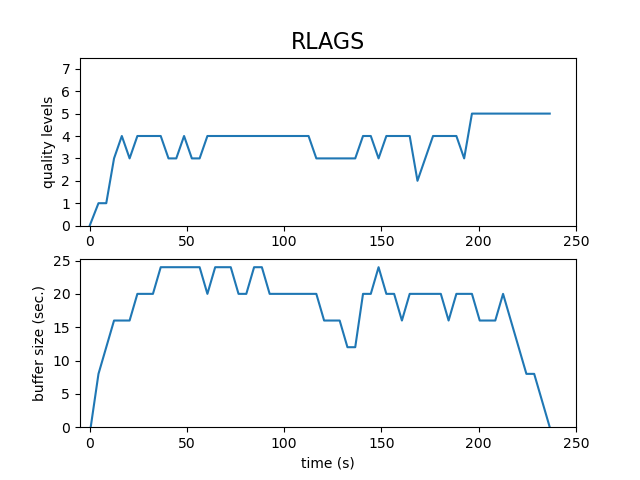} 	
		\includegraphics[width=0.45\textwidth]{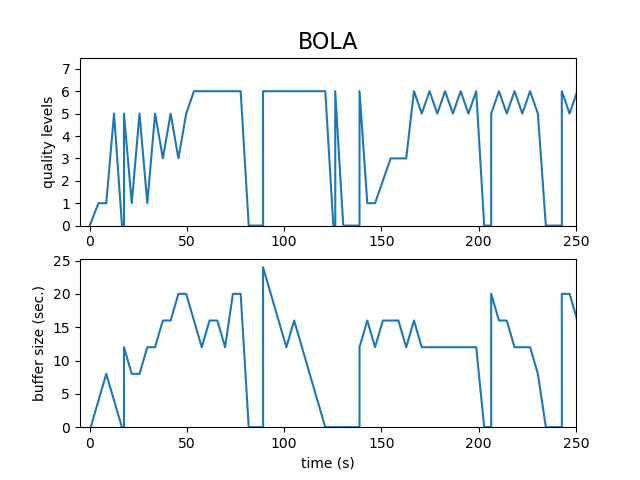}\\
		\includegraphics[width=0.45\textwidth]{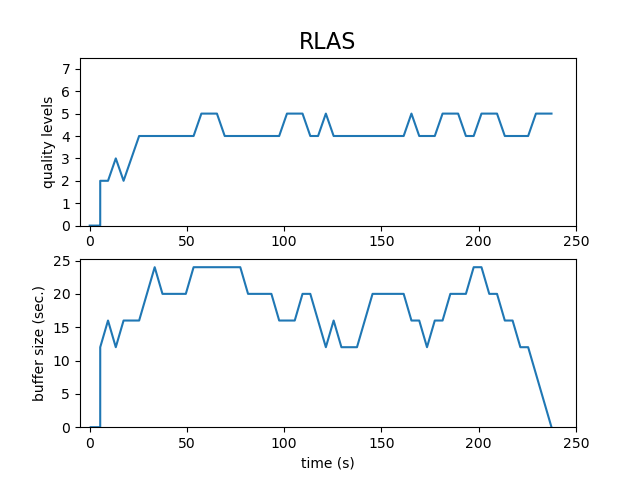}
		\includegraphics[width=0.45\textwidth]{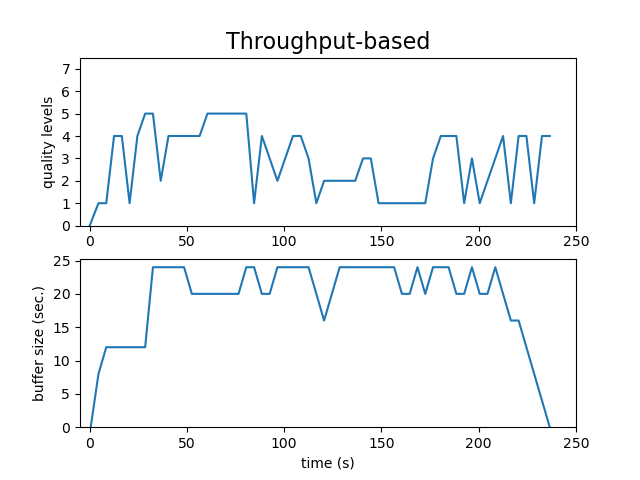}
	\caption{The quality levels and the buffer occupancy with the sample throughput traces in Fig. \ref{fig:bwtraces}.}
	\label{fig:qualityLevelFccLte}	
\end{figure}

Fig.~\ref{fig:qualityLevelFccLte} shows video quality levels selection and buffer occupancy when the client experiences a pair of throughput traces, shown in Fig. \ref{fig:bwtraces}, with different methods. 
The video played by the RL-based methods is more stable than by the other methods. We see that the proposed methods have a smarter buffer occupancy so that they can download higher quality levels with fewer switches than other methods.

\section{Conclusions}
We have proposed two novel adaptation and scheduling methods for video streaming from multiple sources, \textit{i.e.}, RL-based adaptation and greedy scheduling (RLAGS) and RL-based adaptation and scheduling (RLAS).
The state space, action space, and reward are defined for the methods.
We have also built a GymAI-compatible environment for training and evaluating.
Extensive simulations have shown that the proposed methods outperform the baseline methods in terms of the user's QoE.
Model-free reinforcement learning algorithms could not work well in transfer learning \cite{model-based}. If we run the model in an untrained environment, the model could yield a low reward. In the future, we will apply model-based algorithms to bitrate adaptation.

\end{document}